\documentclass[traditabstract,letter]{aa}

\usepackage{graphicx}
\usepackage{natbib}
\usepackage{url}
\usepackage{multirow}
\usepackage{rotating}
\usepackage{txfonts}              
\begin{document}
\defcitealias{Noterdaeme09dla}{N09}

%USER DEFINED ShORTCUTS
\newcommand{\zabs}{\ensuremath{z_{\rm abs}}}
\newcommand{\zmin}{\ensuremath{z_{\rm min}}}
\newcommand{\zmax}{\ensuremath{z_{\rm max}}}
\newcommand{\zem}{\ensuremath{z_{\rm em}}}
\newcommand{\zqso}{\ensuremath{z_{\rm QSO}}}
\newcommand{\dla}{damped Lyman-$\alpha$}
\newcommand{\Dla}{damped Lyman-$\alpha$}
\newcommand{\lya}{Ly-$\alpha$}
\newcommand{\lyb}{Ly-$\beta$}
\newcommand{\lyg}{Ly-$\gamma$}

%ions A\&A style 
\newcommand{\ArI}{\ion{Ar}{i}}
\newcommand{\CaII}{\ion{Ca}{ii}}
\newcommand{\CI}{\ion{C}{i}}
\newcommand{\CII}{\ion{C}{ii}}
\newcommand{\CIV}{\ion{C}{iv}}
\newcommand{\ClI}{\ion{Cl}{i}}
\newcommand{\ClII}{\ion{Cl}{ii}}
\newcommand{\CrII}{\ion{Cr}{ii}}
\newcommand{\CuII}{\ion{Cu}{ii}}
\newcommand{\DI}{\ion{D}{i}}
\newcommand{\FeI}{\ion{Fe}{i}}
\newcommand{\FeII}{\ion{Fe}{ii}}
\newcommand{\HI}{\ion{H}{i}}
\newcommand{\MgI}{\ion{Mg}{i}}
\newcommand{\MgII}{\ion{Mg}{ii}}
\newcommand{\MnII}{\ion{Mn}{ii}}
\newcommand{\NI}{\ion{N}{i}}
\newcommand{\NaI}{\ion{Na}{i}}
\newcommand{\NII}{\ion{N}{ii}}
\newcommand{\NV}{\ion{N}{v}}
\newcommand{\NiII}{\ion{Ni}{ii}}
\newcommand{\OI}{\ion{O}{i}}
\newcommand{\OII}{\ion{O}{ii}}
\newcommand{\OIII}{\ion{O}{iii}}
\newcommand{\OVI}{\ion{O}{vi}}
\newcommand{\PII}{\ion{P}{ii}}
\newcommand{\PbII}{\ion{Pb}{ii}}
\newcommand{\SI}{\ion{S}{i}}
\newcommand{\SII}{\ion{S}{ii}}
\newcommand{\SiII}{\ion{Si}{ii}}
\newcommand{\SiIV}{\ion{Si}{iv}}
\newcommand{\TiII}{\ion{Ti}{ii}}
\newcommand{\ZnII}{\ion{Zn}{ii}}
\newcommand{\AlII}{\ion{Al}{ii}}
\newcommand{\AlIII}{\ion{Al}{iii}}
%Other shortcuts.
\newcommand{\Ho}{\mbox{H$_0$}}
\newcommand{\angstrom}{\mbox{{\rm \AA}}}
\newcommand{\abs}[1]{\left| #1 \right|} % for absolute value
\newcommand{\avg}[1]{\left< #1 \right>} % for average
\newcommand{\kms}{\ensuremath{{\rm km\,s^{-1}}}}
\newcommand{\cmsq}{\ensuremath{{\rm cm}^{-2}}}
\newcommand{\fhi}{\ensuremath{f(N_{\rm HI},\chi)}}
\newcommand{\omegagdla}{\ensuremath{\Omega_{\rm g}^{\rm DLA}}}
\newcommand{\gz}{\ensuremath{g(z)}}

%Institutes
\newcommand{\iap}{Institut d'Astrophysique de Paris, CNRS-UPMC, UMR7095, 98bis bd Arago, 75014 Paris, France\label{iap}}
\newcommand{\uchile}{Departamento de Astronom\'ia, Universidad de Chile, Casilla 36-D, Santiago, Chile\label{uchile}}
\newcommand{\apc}{APC, CNRS UMR 7164, 10 rue Alice Domon et L\'eonie Duquet, 75205 Paris Cedex 13, France\label{apc}}
\newcommand{\princeton}{Princeton University Observatory, Peyton Hall, Princeton, NJ 08544, USA}
\newcommand{\berkeley}{Lawrence Berkeley National Lab, 1 Cyclotron Rd, Berkeley CA, 94720, USA\label{berkeley}}
\newcommand{\zurich}{Institute of Theoretical Physics, University of Zurich, 8057 Zurich, Switzerland\label{zurich}}
\newcommand{\ohio}{Department of Astroomy, the Ohio State University, 140 W. 18th Avenue, Columbus OH 43210-1173, USA}
\newcommand{\portsmouth}{Institute of Cosmology and Gravitation, University of Portsmouth, UK \label{portsmouth}}
\newcommand{\wyoming}{Department of Physics and Astronomy, University of Wyoming, Laramie, WY 82071, USA\label{wyoming}}
\newcommand{\pennsya}{Department of Astronomy and Astrophysics, The Pennsylvania State University, University Park, PA 16802, USA \label{pennsya}}
\newcommand{\pennsyb}{Institute for Gravitation and the Cosmos, The Pennsylvania State University, University Park, PA 16802, USA \label{pennsyb}}
\newcommand{\florida}{\mbox{Astronomy Department, University of Florida, 211 Bryant Space Science Center, PO\,Box\,112055, Gainesville, FL\,32611-2055, USA}\label{florida}}
\newcommand{\bcn}{Institut de Ci\`encies del Cosmos, Universitat de Barcelona, Facultat de F\'isica, Barcelona, Spain\label{bcn}}
\newcommand{\apo}{Apache Point Observatory, P.O. Box 59, Sunspot, NM 88349-0059, USA\label{apo}}
\newcommand{\chicagoa}{Department of Astronomy and Astrophysics, University of Chicago, 5640 South Ellis Avenue, Chicago, IL 60637, USA \label{chicagoa}}
\newcommand{\chicagob}{Enrico Fermi Institute, University of Chicago, 5640 South Ellis Avenue, Chicago, IL 60637, USA \label{chicagob}}

\newcommand{\boss}{BOSS collaboration.}

%___________________________________________________________________________________________________________
\title{Column density distribution and cosmological mass density of neutral gas: Sloan Digital Sky Survey-III Data Release 9 
\thanks{The catalogue is only available in electronic form at the CDS via anonymous ftp to \texttt{cdsarc.u-strasbg.fr} 
\texttt{(130.79.128.5)} or via \texttt{http://cdsweb.u-strasbg.fr/cgi-bin/xxxxxxxxxxxxxxxxxx}.}
}

\titlerunning{Distribution and evolution of neutral gas at $2<z<3.5$ from SDSS DR9}

\author{
            P.~Noterdaeme       \inst{\ref{iap}}
    \and    P.~Petitjean        \inst{\ref{iap}}
    \and W.~C.~Carithers        \inst{\ref{berkeley}}
    \and    I.~P\^aris          \inst{\ref{uchile}}
    \and    A.~Font-Ribera      \inst{ \ref{zurich}, \ref{berkeley}}
    \and    S.~Bailey           \inst{\ref{berkeley}}
    \and    E.~Aubourg          \inst{\ref{apc}}  
    \and    D.~Bizyaev          \inst{\ref{apo}} 
    \and    G.~Ebelke           \inst{\ref{apo}}
    \and    H.~Finley           \inst{\ref{iap}}
    \and    J.~Ge               \inst{\ref{florida}}
    \and    E.~Malanushenko     \inst{\ref{apo}}
    \and    V.~Malanushenko     \inst{\ref{apo}}
    \and    J.~Miralda-Escud\'e \inst{\ref{bcn}}
    \and A.~D.~Myers            \inst{\ref{wyoming}}     
    \and    D.~Oravetz          \inst{\ref{apo}}   
    \and    K.~Pan              \inst{\ref{apo}}   
    \and M.~M.~Pieri            \inst{\ref{portsmouth}}
    \and N.~P.~Ross             \inst{\ref{berkeley}}
    \and D.~P.~Schneider        \inst{\ref{pennsya}, \ref{pennsyb}}
    \and    A.~Simmons          \inst{\ref{apo}}
    \and D.~G.~York             \inst{\ref{chicagoa}, \ref{chicagob}}
}
  
\institute{     \iap\  -- \email{noterdaeme@iap.fr}
           \and
        	\berkeley\ 
           \and
                \uchile\ 
           \and
         	\zurich\ 
           \and 
                \apc\ 
           \and 
                \apo\ 
           \and 
                \florida\ 
           \and 
                \bcn\ 
           \and 
                \wyoming 
           \and
                \portsmouth\ 
           \and 
                \pennsya\ 
           \and 
                \pennsyb\         
           \and 
                \chicagoa\ 
           \and 
                \chicagob\ 
         }

\date{}

\abstract{
We present the first results from an ongoing survey for Damped Lyman-$\alpha$ systems (DLAs)
in the spectra of $z>2$ quasars observed in the course of the Baryon Oscillation Spectroscopic Survey 
(BOSS), which is part of the Sloan Digital Sky Survey (SDSS) III. Our full (non-statistical) sample, 
based on Data Release 9, comprises {12,081} systems with $\log N(\HI) \ge 20$, out of which 
{6,839} have $\log N(\HI)\ge 20.3$. This is the largest DLA sample ever compiled, superseding that 
from SDSS-II by a factor of seven. 

Using a statistical sub-sample and estimating systematics from realistic mock data, 
we probe the $N(\HI)$ distribution at $\avg{z} = 2.5$.  Contrary 
to what is generally believed, the distribution extends beyond 10$^{22}$~cm$^{-2}$ with 
a moderate slope of index $\approx -3.5$.
This result matches surprisingly well 
the opacity-corrected distribution observed at $z=0$.
The cosmological mass density of neutral gas in DLAs is found to be $\omegagdla \approx 10^{-3}$, 
evolving only mildly over the past 12 billion years.
}

\keywords{cosmology: observations - quasar: absorption-lines - galaxies:evolution}

\maketitle
%___________________________________________________________________________________________________________

\section{Introduction}

Studying the distribution of neutral gas in and around galaxies at different cosmological times 
provides a wealth of information about the formation and evolution of  galaxies. 
The 21-cm hyperfine emission of atomic hydrogen has been used to trace the neutral gas in nearby
galaxies and estimate their total \HI\ mass. Given the sensitivity of present day 
radio telescopes,  this technique remains limited to $z \le 0.2$  \citep[e.g.][]{Lah07}. 
At high redshift, neutral gas is revealed by the damped Lyman-$\alpha$ absorption systems (DLAs) it
imprints in the optical spectra of bright background sources such as quasars. 
Because the detection of DLAs is only cross-section dependent, it is 
possible to statistically derive the amount of neutral gas and the corresponding column density distribution 
at different redshifts independently of the nature of the absorbers 
\citep[see][]{Wolfe05}.

The most recent contributions to the census of DLAs used data mining of 
thousands of quasar spectra from the Sloan Digital Sky Survey \citep[SDSS,][]{York00}  
by \citet[][hereafter \citetalias{Noterdaeme09dla}]{Prochaska05,Prochaska09a,Noterdaeme09dla}. 
These studies indicate that the $N(\HI)$ distribution function (\fhi{, where $\chi$ is the absorption 
distance, see \citealt{Lanzetta91}}) steepens 
at $\log N(\HI) > 21$ and that the cosmological density of 
neutral gas contained in DLAs ($\omegagdla$) decreases significantly 
with time between $z \sim 3.5$ and $z=2.2$.

Several explanations for the steepening of \fhi\ have been discussed in the literature, 
including conversion from atomic to molecular hydrogen \citep{Zwaan06}, small-scale 
turbulence, or stellar feedback \citep{Erkal12}. Selection effects such as dust-reddening 
\citep[e.g.][]{Vladilo05} could also alter the slope of \fhi\ in magnitude-limited surveys. 
However, the slope of the frequency distribution itself is not yet well constrained at the 
high-column-density end due to rapidly decreasing statistics. 
Similarly, the evolution of $\omegagdla$ has been long discussed in the literature. 
Values at $z \sim 1$ \citep{Rao06} have been considered uncomfortably high when compared to 
that at $z=0$ and $z \sim 2$ from \citet{Zwaan05} and \citet{Prochaska05} respectively. 
However, \citetalias{Noterdaeme09dla} corrected upwards the value at $z \sim 2$. 
Since then, the value at $z=0$ has also been corrected upwards by \citet{Braun12} and could 
indicate a flatter evolution over $0<z<2$.

In this letter we present a search for DLAs in quasars observed in the course of the Baryonic 
Oscillation Spectroscopic Survey \citep[BOSS,][]{Dawson12}, one of the legacy surveys 
in the third stage of the SDSS \citep{Eisenstein11}.
We use the same formalism as described in \citetalias{Noterdaeme09dla} and adopt a $\Lambda$CDM 
cosmology with $\Omega_{\Lambda}=0.73$, $\Omega_{\rm m}=0.27$, and $\Ho=70$~\kms\,Mpc$^{-1}$ \citep[][]{Komatsu11}.

\section{Method \label{qs}}

BOSS is a five-year program using improved spectrographs \citep{Smee12} on the SDSS telescope \citep{Gunn06} 
to obtain spectra of 1.5 million galaxies and over 150,000 $z>2.15$ quasars reaching up to 1\,mag deeper than SDSS-II.
The survey is mainly designed to measure the characteristic scale imprinted by baryon acoustic oscillations 
(BAOs) in the early Universe from the spatial distribution of luminous galaxies at $z\sim 0.7$ and 
the large-scale correlation of \HI\ absorption lines in the intergalactic medium at $z\sim 2.5$ \citep{Dawson12}. 
BOSS uses the same imaging data as in SDSS-I and II with an extension in the south galactic cap \citep[see][]{DR8}. 
The SDSS-DR9 \citep{Ahn12} makes publicly available the spectra of 87,822 quasars over an area of 
3,275 ${\rm deg^2}$, 65,205 having $z>2$ \citep{Paris12}. 
The quasar target selection is described in \citet[][see also \citealt{Bovy11}]{Ross12}.

\subsection{Detection of DLAs}
Intervening DLAs were searched for automatically in quasar spectra following the method described 
in \citetalias{Noterdaeme09dla}. 
We briefly summarise here the main steps. For the purpose of collecting the 
largest number of DLAs\footnote{DLAs are contaminants for the study of the \lya\ 
forest correlation function \citep{Font-Ribera12b}.}, we searched the full line-of-sight to 
each quasar starting where the spectral signal-to-noise ratio per pixel reaches 2 (defining 
$\zmin$) and up to the quasar redshift. 
We avoid sight-lines with broad absorption lines with balnicity index BI$>$1000~\kms \citep{Paris12}. 

The quasar continuum is modelled over the \lya\ forest by fitting a modified power-law with a smoothly 
changing index plus Moffat profiles on top of the emission lines. Whenever the \lya\ 
emission line was severely absorbed ($>$30\%), we used the predicted 
unabsorbed emission from principal component analysis \citep[see][]{Paris11} as a proxy for the 
true \lya\ emission before fitting the continuum. 
We then use the median continuum-to-noise ratio as an estimate of the quality of the spectrum, 
independent of the presence of a DLA. Spectra with median CNR~$<$~2 over the \lya\ forest were 
not further considered.

Damped absorption lines are recognised through their characteristic shape by correlating the data against 
synthetic profiles of increasing column density \citepalias[see][]{Noterdaeme09dla}. In short, ($N(\HI),z$) 
pairs with Spearman's correlation above 0.5 
(and significance $>3$\,$\sigma$) are recorded. To constrain the strength of the absorption, we also impose 
that the absorbed flux should be consistent with the presence of a DLA combined with possible \lya\ forest 
absorptions. The pairs 
are then grouped into individual DLA candidates (a gap of $>$1000~\kms\ indicates separate absorption 
systems), and the first guess for $N(\HI)$ is taken from the pair with the highest correlation. 
The DLA redshift measurement is then improved whenever possible 
by cross-correlating the QSO spectrum on the red side of the Lyman-$\alpha$ emission line with a mask 
representing metal absorption 
lines. Finally, $N(\HI)$ is obtained by fitting a Voigt profile to the \dla\ line. 

This approach provides us with an overall sample of 12,081 DLA candidates with $\log N(\HI) \ge 20$, out 
of which 6,839 have $\log N(\HI) \ge 20.3$ (Table~3, in the electronic version only). We also provide 
values of (or limits on) 
the equivalent widths of associated metal lines redwards of the \lya\ emission line. 

\subsection{Statistical sample}
We subsequently define a {\sl statistical} sub-sample that is used to derive the $N(\HI)$ distribution 
function and the integrated cosmological mass density of neutral gas. 
First of all, we conservatively reject all QSOs with even moderate balnicity 
\citep[BI$>0$ or flagged visually][]{Paris12} and 
apply a more stringent threshold on the data quality, keeping only spectra with {CNR~$>3$}.
We then restrict the redshift range as follows: i) \citetalias{Noterdaeme09dla} showed that the 
presence of a DLA near the blue end of the spectrum can bias the definition of $\zmin$ and proposed 
a systematic 10000~\kms\ velocity shift to $\zmin$ that we also apply here. 
ii) We consider only the region 3000~\kms redwards of the \lyb\ emission line and 
5000~\kms\ bluewards of the \lya\ emission line. The 
first cut ensures that we consider only the \lya\ forest and avoid the \lyb+\OVI\ region where associated 
broad \OVI\ absorption can occur (even if no broad \CIV\ absorption is seen) and be mistaken as DLAs. 
The second cut avoids DLAs located in the vicinity of the QSO \citep[e.g.][]{Ellison02}. 
Finally, we restrict our study to the range $z\in[2,3.5]$. This avoids the very blue 
end of the spectra (below 3650~{\AA}) where reduction problems have been identified \citep{Paris12}.
We set the upper limit because the 
increasing density of the \lya\ forest can introduce a significant fraction of false DLA identifications 
due to strong blending of the \lya\ forest lines at the SDSS resolution \citep{Rafelski12}. The value of 3.5 
corresponds to the redshift out to which this systematic can be reliably estimated based on an analysis of mock 
spectra. 

\begin{figure}
\centering
\includegraphics[bb = 88 187 480 383,clip=,width=0.8\hsize]{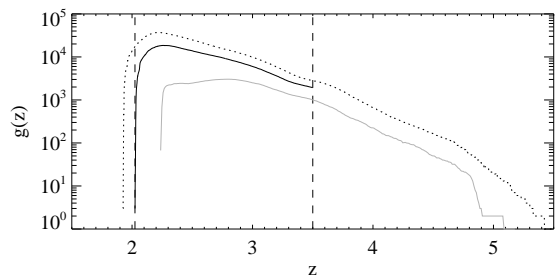}
\caption{Redshift sensitivity function $\gz$ of our full DR9 sample (dotted) and statistical sample (black) 
compared to that of DR7 (\citetalias{Noterdaeme09dla}, grey). 
\label{gz}}
\end{figure}

These cuts leave us with {37,503} lines-of-sight bearing {5,428} systems with $\log N(\HI) \ge 20$ 
({3,408} bona-fide DLAs with $\log N(\HI) \ge 20.3$). We present in Fig.~\ref{gz} the sensitivity functions 
$\gz$ (i.e. the number of lines-of-sight covering a given redshift) for the full and statistical samples. 
Our statistical DR9 sample is 
more than 3 times larger than the overall DR7 sample and also extends to lower redshifts, thanks to the improved blue coverage 
of the SDSS spectrograph. The total absorption path length probed by our statistical sample over $z=2-3.5$ 
is $\Delta \chi \approx$~{45,000} with an average redshift $\avg{z} = 2.5$.

\subsection{Mocks}
The BOSS collaboration is constantly developing mocks that simulate the Lyman-$\alpha$ forest 
seen towards BOSS QSOs \citep[e.g.][]{Font-Ribera12a}. While mocks were principally 
designed for BAO studies, the important point for this study is that simulated spectra are produced 
with the same noise and flux distributions as in the actual DR9 data (Bailey et al., in prep.). 
Furthermore, DLAs have been introduced to the mocks with a known distribution \citep{Font-Ribera12b}. 
We also applied our DLA-searching algorithm to 33 realisations of 3,861 mocks representative of the 
DR9 data with the same cuts as in real data.
From this exercise, the completeness and purity (1 minus the fraction of false identifications) in the 
statistical sample are both found to be above 95\% for $\log N(\HI) \ge 20.3$ (and higher 
when restricting to higher $N(\HI)$ systems). Overall, the
automatic procedure systematically overestimates $N(\HI)$ by 0.03~dex. This is much 
lower than the dispersion (0.20~dex) which corresponds to the typical 1\,$\sigma$ error on $\log N(\HI)$.

\section{The column density distribution at $\avg{z}=2.5$}

\begin{figure*}%[!ht]
\centering
\renewcommand{\tabcolsep}{-0.5pt}
\begin{tabular}{ccc}
\includegraphics[bb = 76 178 494 570,clip=,height=0.31\hsize]{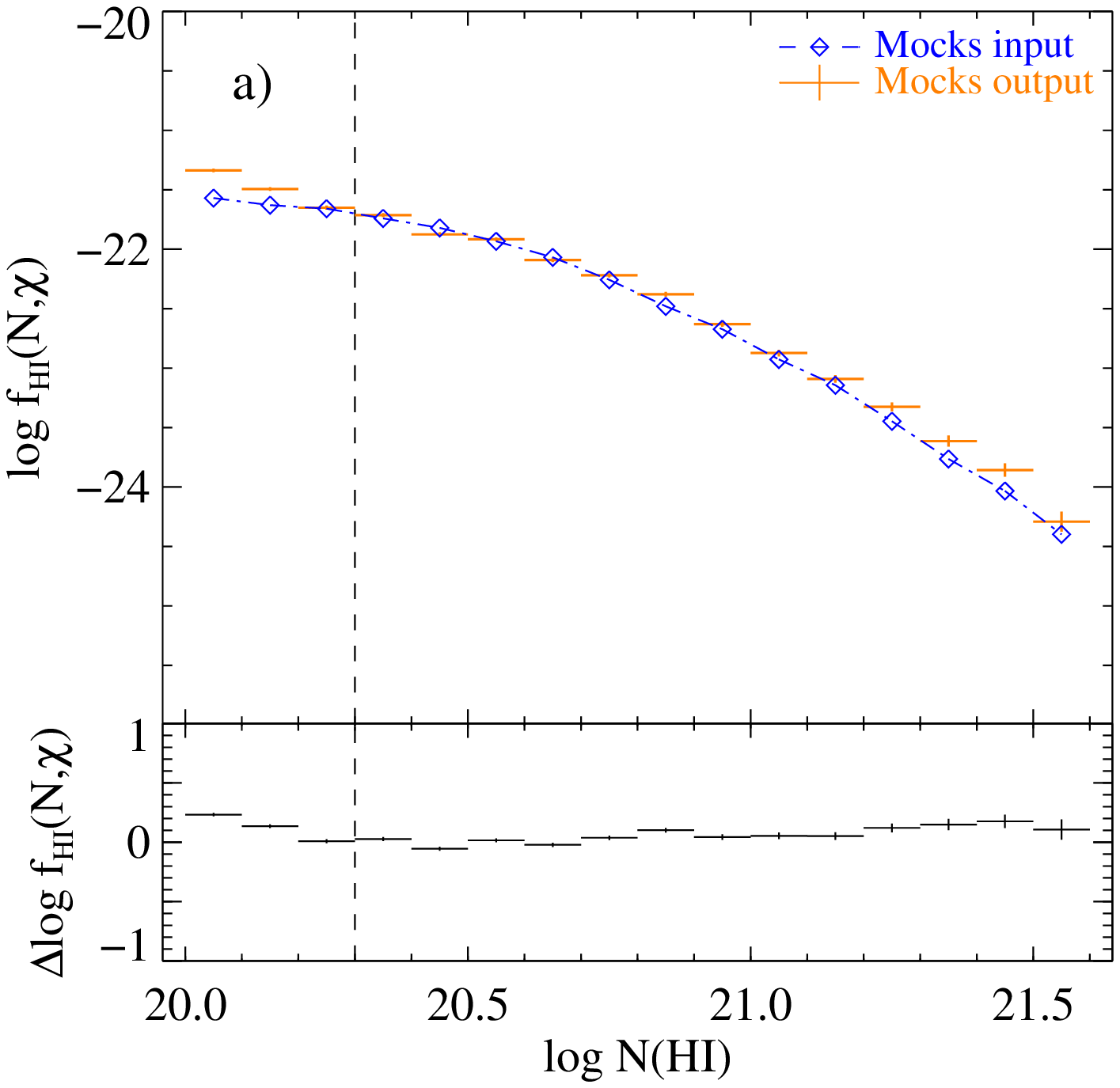}&
\includegraphics[bb = 76 178 494 570,clip=,height=0.31\hsize]{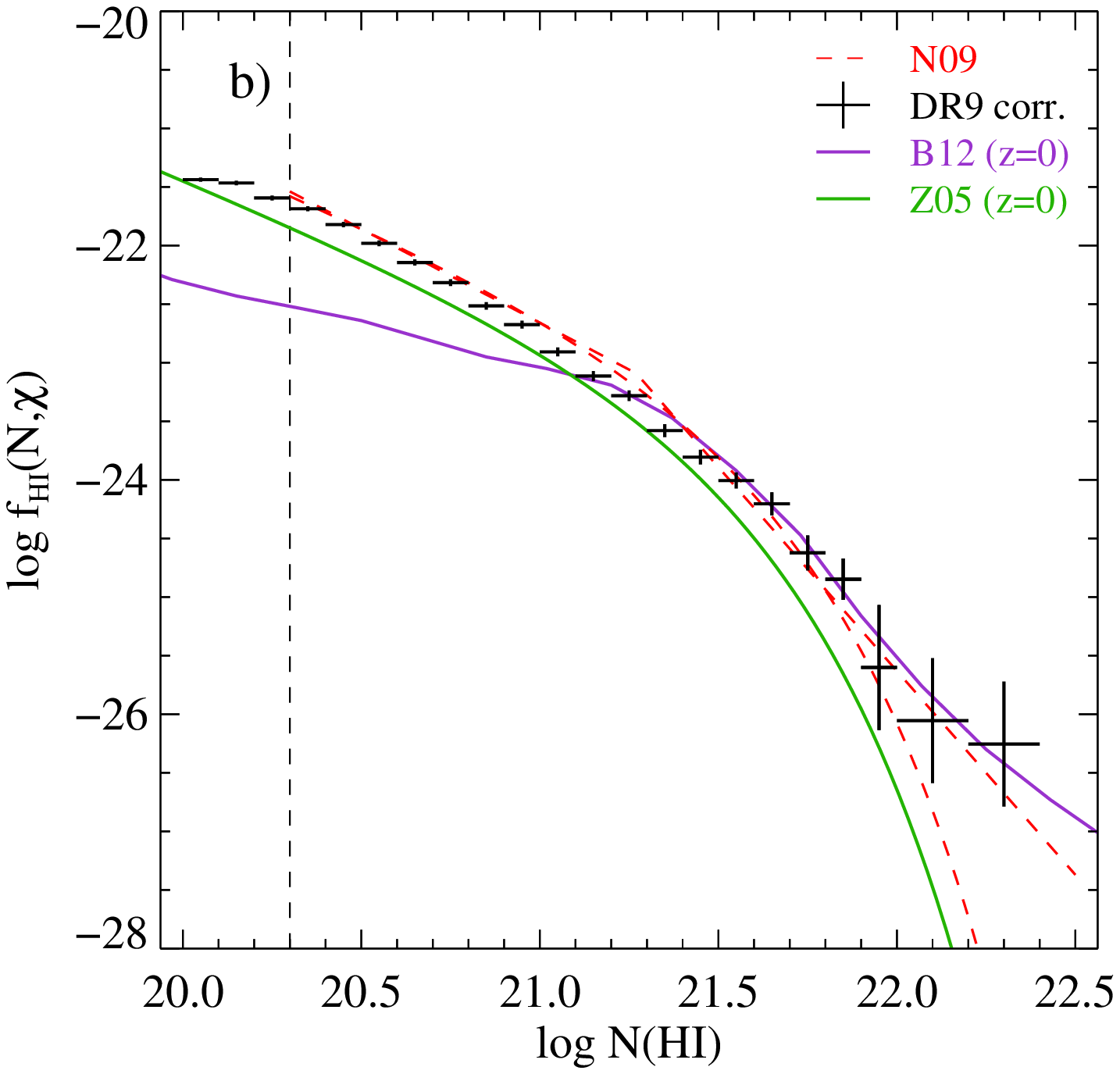}&
\includegraphics[bb = 76 178 520 570,clip=,height=0.31\hsize]{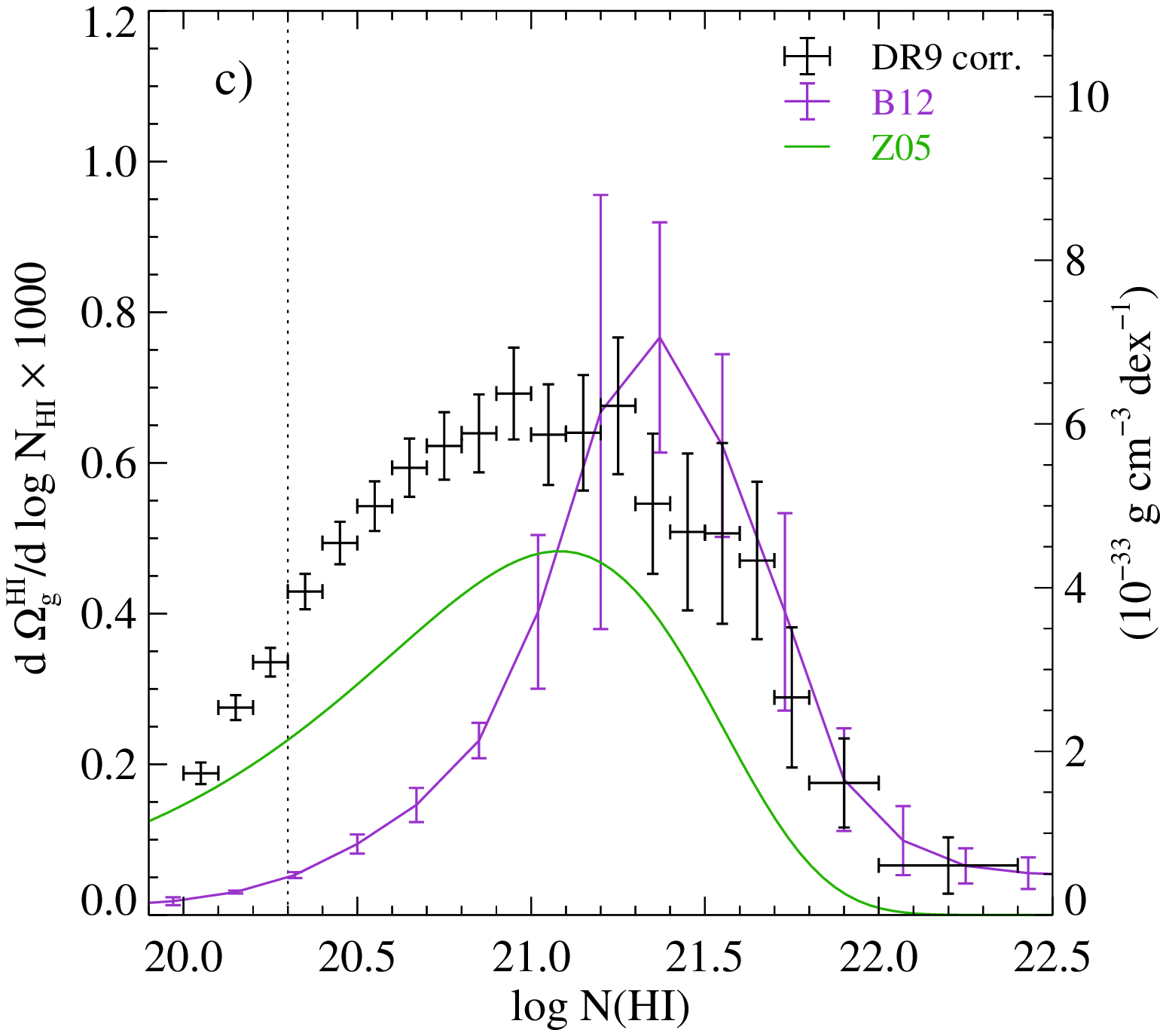}\\
\end{tabular}
\renewcommand{\tabcolsep}{6pt}
\caption{Column density distribution functions from synthetic (left) and real data (centre) at $\avg{z}=2.5$. 
Horizontal bars 
represent the bin over which \fhi\ is calculated and vertical error bars represent Poissonian uncertainty. 
The difference between output and input mock distributions is shown at the bottom of panel {\sl a}. 
The double power-law and $\Gamma$-function fits to the DR7 distribution (\citetalias{Noterdaeme09dla}, 
$\avg{z}=2.9$) are shown as red dashed lines. $\fhi(z=0)$ are taken from \citet[][purple]{Braun12} and 
\citet[][green]{Zwaan05}. 
Right: The contribution of DLAs in a given $N(\HI)$ range to the total mass census of neutral gas. 
DR9 values are corrected for systematics. \label{distrib}}
\end{figure*}

In Fig.~\ref{distrib}a, we compare the simulated input distribution of \HI\ column densities (in the range 
$N(\HI) = 10^{20}$ - $4\times10^{21}$~\cmsq) at $z=$~2-3 with that recovered from mocks by our procedure over the 
same redshift range. 
We can see that the overall agreement is excellent; although our procedure slightly overestimates \fhi, 
particularly at the low column density end. 
We use the difference between the input and output distributions 
as the correction to apply to the observed distribution from real data. 

To ascertain the properties of the high-column-density end of \fhi\ --where statistics are much smaller--  
we have visually checked all DLA candidates with $\log N(\HI) \ge 21.6$. In this 
regime, blind correction using mocks could be more uncertain as the corresponding \HI\ fits are based on \lya\ 
only while metals are systematically detected in the real data.
Indeed, we found a few cases where two closely-spaced DLAs were mistaken for a higher column density one. 
Disentangling such blends was possible thanks to the presence of metal lines. 
For each DLA candidate with $\log N(\HI) \ge 21.6$, the absorption profile was carefully refitted 
manually, improving the continuum determination and using metal lines to determine a precise redshift of the 
absorber. 
The resulting \fhi\ at $z=2-3$ is shown in Fig.~\ref{distrib}b with values given in Table~\ref{tab:fhi}.
It is apparent that the distribution extends beyond 10$^{22}$~cm$^{-2}$ with 5 systems with 
$\log N(\HI) \ge 22$ in the statistical sample (8 in the full sample). 
Extrapolating this function, we might expect to detect DLAs reaching $\log N(\HI)=23$ at the completion 
of BOSS. 

Following \citetalias{Noterdaeme09dla}, 
we measure the total amount of neutral gas in DLAs at $\avg{z}=2.5$ to be $\omegagdla \approx 10^{-3}$. 
Fig.~\ref{distrib}c represents the contribution to the total amount of neutral gas as a function of $N(\HI)$. 
We confirm \citetalias{Noterdaeme09dla}'s result that the largest contribution comes from systems with 
$N(\HI)\sim 10^{21}$\,\cmsq. However, it is interesting that the systems with $N(\HI)$ in excess of 
$5 \times 10^{21}$~\cmsq\ contribute a non-negligible fraction of $\omegagdla$ ($\sim$ 10\%), although they are rarely 
represented in most surveys.

\begin{table}
\renewcommand{\tabcolsep}{4pt}
\centering
\caption{$N(\HI)$ distribution function at $\avg{z}=2.5$ \label{tab:fhi}}
\begin{tabular}{c c c c c}
\hline
\hline
{\large \strut} $\log N(\HI)$ & $\log \fhi$ & $\log \fhi_{\rm corr.}$ \tablefootmark{a} & $\sigma(\log \fhi)$\tablefootmark{b} \\
\hline
$[$20.00,20.10$[$ & -21.20 & -21.44 & 0.02\\
$[$20.10,20.20$[$ & -21.37 & -21.47 & 0.02\\
$[$20.20,20.30$[$ & -21.55 & -21.59 & 0.02\\
$[$20.30,20.40$[$ & -21.66 & -21.68 & 0.02\\
$[$20.40,20.50$[$ & -21.81 & -21.82 & 0.02\\
$[$20.50,20.60$[$ & -21.97 & -21.98 & 0.02\\
$[$20.60,20.70$[$ & -22.13 & -22.14 & 0.03\\
$[$20.70,20.80$[$ & -22.30 & -22.32 & 0.03\\
$[$20.80,20.90$[$ & -22.49 & -22.51 & 0.03\\
$[$20.90,21.00$[$ & -22.63 & -22.67 & 0.03\\
$[$21.00,21.10$[$ & -22.85 & -22.91 & 0.04\\
$[$21.10,21.20$[$ & -23.04 & -23.11 & 0.04\\
$[$21.20,21.30$[$ & -23.19 & -23.28 & 0.05\\
$[$21.30,21.40$[$ & -23.46 & -23.58 & 0.06\\
$[$21.40,21.50$[$ & -23.66 & -23.81 & 0.07\\
$[$21.50,21.60$[$ & -23.83 & -24.01 & 0.08\\
$[$21.60,21.70$[$ & -24.20 & -24.20 & 0.08\\
$[$21.70,21.80$[$ & -24.62 & -24.62 & 0.12\\
$[$21.80,21.90$[$ & -24.85 & -24.85 & 0.18\\
$[$21.90,22.00$[$ & -25.60 & -25.60 & 0.53\\
$[$22.00,22.20$[$ & -26.05 & -26.05 & 0.53\\
$[$22.20,22.40$[$ & -26.25 & -26.25 & 0.53\\
\hline
\end{tabular}
\tablefoot{
\tablefoottext{a}{Corrected for systematics.}
\tablefoottext{b}{Poissonian errors.}
}
\end{table}

\section{Cosmological mass density of neutral gas} 

\begin{figure}
\centering
\includegraphics[bb = 65 180 500 570,clip=,width=0.83\hsize]{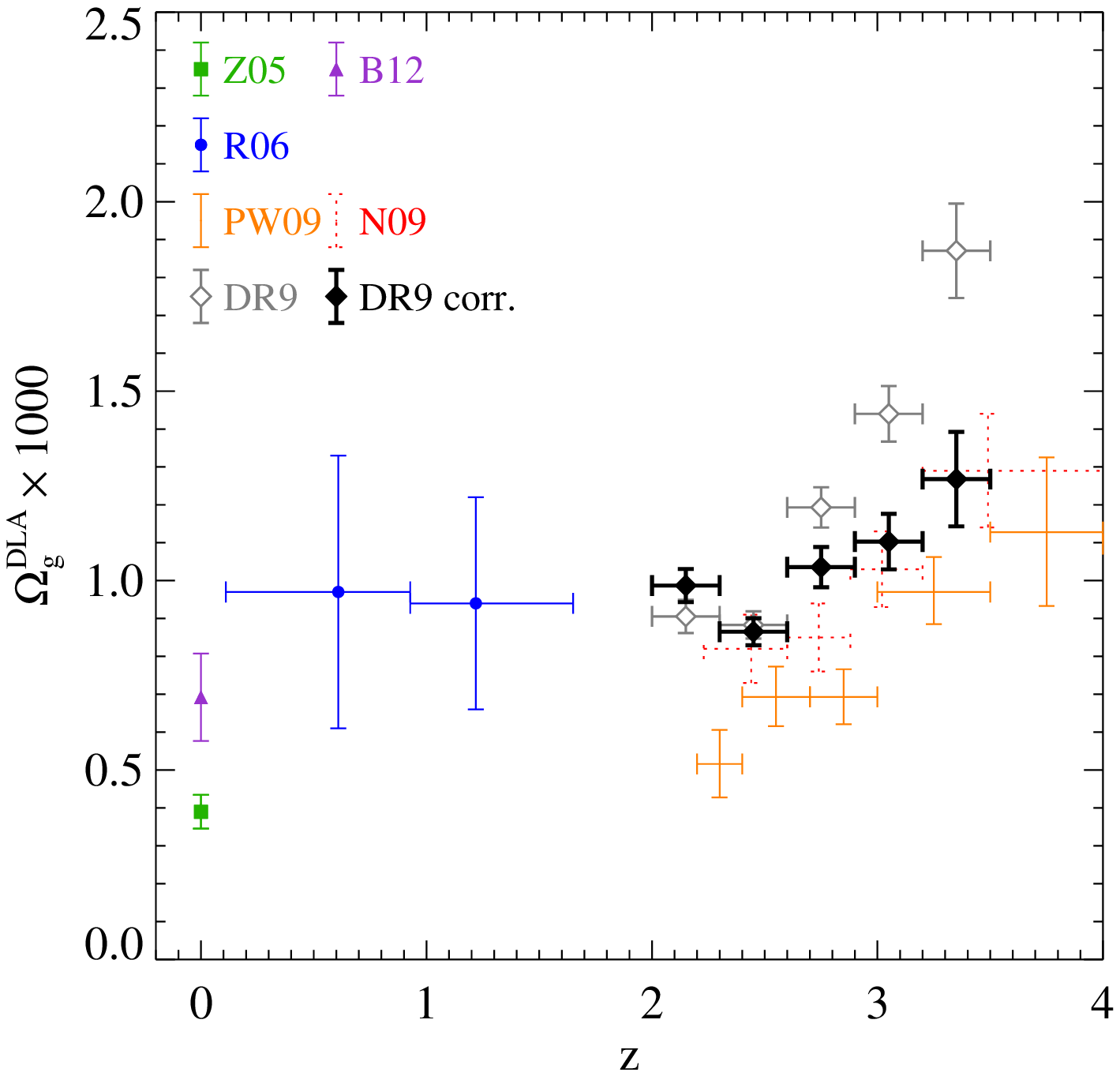} %\\
\caption{Cosmological mass density of neutral gas in DLAs as a function of redshift 
(Z05: \citet{Zwaan05}, B12: \citet{Braun12}, R06: \citet{Rao06}, PW09: \citet{Prochaska09a}, 
DR9: this work). \label{omegaf}}
\end{figure}

Fig.~\ref{omegaf} (see also Table~\ref{omegat}) shows the evolution of the cosmological mass density in 
DLAs as a function of 
redshift. Using mock spectra, we estimate a correction for systematics 
(over/under-estimate of $N(\HI)$, incompleteness and contribution of false positives) as a function 
of redshift. At high redshift, the correction is mostly due to $N(\HI)$ overestimation due to the denser 
\lya\ forest together with increasing false positive identifications.
At $z<2.3$, the correction is upwards due to higher incompleteness and slight underestimation 
of $N(\HI)$. 
Note that the zero-point photometric calibration can be in error by 
about 5\% below 4,000~{\AA} \citep{Paris12}, which could differently affect the detection of DLAs and 
$N(\HI)$-measurements in mocks and real data. This problem will be addressed by forthcoming versions 
of the pipeline \citep{Bolton12}.

We observe a decrease of $\omegagdla$ from $z=3.5$ to $z=2.3$ as in \citetalias{Noterdaeme09dla} 
and \citet{Prochaska09a}, although with higher values at $z<3.2$. 
This can be explained by the $\sim$10\% contribution of very large column density systems in DR9 and 
better knowledge of systematics.
It is unclear which value of $\omegagdla$ should be used at $z=0$. The measurement by \citet{Braun12} is 
based on only three galaxies and although the high-column density end of \fhi\ seems to be well constrained, 
this may not be true at low $N($H\,{\sc i}) \citep{Zwaan05}.
Measurements at $z \sim 1$ \citep{Rao06} are still indirect and, while direct searches for 
DLAs at low redshift are possible \citep{Meiring11}, they are still quite limited in terms of sample size.
It appears that systematics dominate over statistical uncertainties across most of the redshift range. 
Keeping this in mind, we can still conclude that $\omegagdla$ evolves only mildly over the past 12 Gyr.

\section{Conclusion}

We have presented the first results of our ongoing survey for DLAs in the SDSS-III Baryon Oscillation 
Spectroscopic Survey, Data Release 9. This represents by far the largest sample of DLAs to date (with 
$\sim$12,000 systems with $\log N(\HI)\ge 20$) and should allow numerous follow-up studies.
We expect the sample to be increased by a factor larger than two at the completion of BOSS.
Using a well defined sub-sample, and controlling systematics (which dominate over statistical errors) 
with synthetic spectra, we derive the \HI\ column density distribution at $\avg{z}=2.5$ in the range 
$10^{20}-2\times10^{22}$~\cmsq\ and characterise the evolution of the cosmological mass density of neutral 
gas in DLAs at $2<z<3.5$. 
This study should help to constrain models of galaxy formation and evolution by measuring the amount of neutral 
gas immediately available to fuel star formation through cosmic history.

\begin{table}
\centering
\renewcommand{\tabcolsep}{3pt}
\caption{$\omegagdla$ and DLA incidence ($dN/dz$) in different redshift bins. \label{omegat}}
\begin{tabular}{c c c c c c}
\hline
\hline
$z$                                                   & 2.0 - 2.3    & 2.3 - 2.6    & 2.6 - 2.9    & 2.9 - 3.2    & 3.2 - 3.5    \\
$\Delta z$                                            & 3690         & 4509         & 2867         & 1620         & 769          \\
$\Delta \chi$~\tablefootmark{\ast}                    & 11625        & 14841        & 9900         & 5834         & 2883         \\  
$10^3$\,\omegagdla~\tablefootmark{\dagger}            &0.91/0.99      &0.88/0.87    & 1.19/1.04    & 1.44/1.10    & 1.87/1.27    \\
$10^3$\,$\sigma(\omegagdla)$\,\tablefootmark{\ddagger}        &0.05          &0.04          & 0.05         & 0.08         & 0.13      \\ 
$dN/dz$~\tablefootmark{\dagger}                                        &0.19/0.20          &0.21/0.20          & 0.29/0.25         & 0.36/0.29         & 0.48/0.36       \\
\hline
\end{tabular}
\tablefoot{
\tablefoottext{\ast}{Total absorption pathlength \citep[see][]{Lanzetta91}.}
\tablefoottext{\dagger}{Direct values/corrected for systematics.}
\tablefoottext{\ddagger}{Statistical uncertainty.}
}
\end{table}

\begin{acknowledgements}
We thank the anonymous referee for helpful comments and suggestions.
The French participation group to SDSS-III was supported by the Agence Nationale de la
Recherche under grant ANR-08-BLAN-0222.
Funding for SDSS-III has been provided by the Alfred P. Sloan Foundation, the Participating 
Institutions, the National Science Foundation, and the U.S. Department of Energy Office of Science. 
The SDSS-III web site is \url{http://www.sdss3.org/}.
SDSS-III is managed by the Astrophysical Research Consortium for the Participating Institutions 
of the SDSS-III Collaboration including the University of Arizona, the Brazilian Participation 
Group, Brookhaven National Laboratory, University of Cambridge, Carnegie Mellon University, 
University of Florida, the French Participation Group, the German Participation Group, Harvard 
University, the Instituto de Astrof\'isica de Canarias, the Michigan State/Notre Dame/JINA Participation 
Group, Johns Hopkins University, Lawrence Berkeley National Laboratory, Max Planck Institute for 
Astrophysics, Max Planck Institute for Extraterrestrial Physics, New Mexico State University, New 
York University, Ohio State University, Pennsylvania State University, University of Portsmouth, 
Princeton University, the Spanish Participation Group, University of Tokyo, University of Utah, 
Vanderbilt University, University of Virginia, University of Washington, and Yale University.

\end{acknowledgements}

\bibliographystyle{aa}

\end{document}